\documentstyle[aps,epsfig,prd]{revtex}

\begin{document}

\twocolumn[\hsize\textwidth\columnwidth\hsize\csname
@twocolumnfalse\endcsname

\title{CMB Bispectrum from Active Models of Structure Formation}

\author{Alejandro Gangui\( ^{1,2} \)\footnotemark[1],
        Levon Pogosian\( ^{3} \)\footnotemark[2],
        and
        Serge Winitzki\( ^{3} \)\footnotemark[3]{} }

\address{\( ^{1} \)Instituto de Astronom\'{\i}a y F\'{\i}sica del Espacio,
Ciudad Universitaria, 1428 Buenos Aires, Argentina}

\address{\( ^{2} \)Dept. de F\'{\i}sica, Universidad de Buenos Aires,
Ciudad Universitaria -- Pab. 1, 1428 Buenos Aires, Argentina}

\address{\( ^{3} \)Physics Department, Case Western Reserve University,
Cleveland, Ohio 44106-7079, USA}

\maketitle

\begin{abstract}
We propose a new method for a numerical computation of the angular bispectrum
of the CMB anisotropies arising from active models such as cosmic topological
defects, using a modified Boltzmann code based on \textsc{CMBFAST}. The method
does not use CMB sky maps and requires moderate computational power. As a first
implementation, we apply our method to a recently proposed model of simulated
cosmic strings and estimate the observability of the non-Gaussian bispectrum
signal. A comparison with the cosmic variance of the bispectrum estimator shows
that the bispectrum for the simulated string model we used is not observable.
\end{abstract}
\vskip.5pc]\footnotetext[1]{gangui@iafe.uba.ar}
\footnotetext[2]{levon@theory6.phys.cwru.edu}
\footnotetext[3]{winitzki@erebus.phys.cwru.edu}

\noindent

Anisotropies of the Cosmic Microwave Background radiation (CMB) are
directly related to the origin of structure in the universe. Galaxies and
clusters of galaxies eventually formed by gravitational instability from
primordial density fluctuations, and these same fluctuations left their
imprint on the CMB. Recent balloon \cite{boomerang,maxima} and 
ground-based interferometer \cite{dasi} experiments have
produced reliable estimates of the power spectrum of the CMB temperature
anisotropies. 
While they helped eliminate certain candidate theories for the
primary source of cosmic perturbations, the power spectrum data is
still compatible with the theoretical estimates of a relatively large
variety of models, such as $\Lambda$CDM, quintessence models or some
hybrid models including cosmic defects.
These models, however, differ in their predictions for
the statistical distribution of the anisotropies beyond the power spectrum.
Future MAP and Planck satellite missions (scheduled for launch in 2001 and
2007, respectively) will provide high-precision data allowing definite
estimates of non-Gaussian signals in the CMB. It is therefore important to
know precisely which are the predictions of all candidate models for the
statistical quantities that will be extracted from the new data and
identify their specific signatures.

There are two main classes of models of structure
formation---\textit{passive} and \textit{active} models. In passive models,
density inhomogeneities are set as initial conditions at some early time,
and while they subsequently evolve as described by Einstein-Boltzmann
equations, no additional perturbations are seeded. On the other hand, in
active models the sources of density perturbations are time-dependent.

All specific realizations of passive models are based on the idea of
inflation. In most inflationary models, density fluctuations arise
from quantum fluctuations of a scalar field placed in the vacuum
and hence are well described by a Gaussian distribution, while
second-order effects may add a weak non-Gaussian signal
\cite{ganguietal94}.

On the other hand, active models of structure formation are
motivated by cosmic topological defects. If our ideas about grand
unification are correct, then some cosmic defects, such as domain walls,
strings, monopoles or textures, should have formed during phase transitions
in the early universe \cite{vilshel}. Once formed, cosmic strings could
survive long enough to seed density perturbations. Defect models possess
the attractive feature that they have no parameter freedom, as all the
necessary information is in principle contained in the underlying particle
physics model. Generically, perturbations produced by active models are not
expected to be Gaussian distributed.

The narrow main peak and the presence of the second and the third 
peaks in the CMB angular power spectrum, as measured by
BOOMERANG, MAXIMA and DASI \cite{boomerang,maxima,dasi}, is 
an evidence for coherent oscillations of the
photon-baryon fluid at the beginning of the decoupling epoch
\cite{agscien}. While such
coherence is a property of all passive model, realistic cosmic string
models produce highly incoherent perturbations that result in a much
broader main peak. This excludes cosmic strings as the primary source of
density fluctuations unless new physics is postulated, \textit{e.g.}~varying
speed of light \cite{varyingc}. In addition to purely active or passive
models, it has been recently suggested that perturbations could be seeded
by some combination of the two mechanisms. For example, cosmic strings
could have formed just before the end of inflation and partially
contributed to seeding density fluctuations. It has been shown
\cite{hybrid} that such hybrid models can be rather successful in fitting
the CMB power spectrum data. Therefore, statistics beyond the
power spectrum is required to discriminate between active and passive
models.

Of the available non-Gaussian statistics, the CMB bispectrum, or the
three-point function of Fourier components of the temperature anisotropy, has
been perhaps the one best studied in the literature \cite{bispvar,GM99}.
Although there are a few cases where the bispectrum may be estimated
analytically from the underlying model, a precise numerical code to compute it,
similar to the \textsc{CMBFAST} code \cite{cmbfast} for the power spectrum, is
presently lacking. The bispectrum can be estimated from simulated CMB sky maps;
however, computing a large number of full-sky maps resulting from defects is a
much more demanding task.

In this article we introduce a new method for obtaining the CMB bispectrum
directly from numerically simulated defect models, without building CMB sky
maps. Given a suitable model, one can generate a statistical
\emph{ensemble} of realizations of defect matter perturbations. We use a
modified Boltzmann code based on \textsc{CMBFAST} to compute the effect of
these perturbations on the CMB and find the bispectrum estimator for a
given realization of sources. We then perform statistical averaging over
the ensemble of realizations to compute the expected CMB bispectrum. (The
CMB power spectrum is also obtained as a byproduct.) Our method is
specifically tailored for computations of the bispectrum; extending it to
higher-order correlation functions would require prohibitively longer
calculations. As a first application, we computed the expected CMB
bispectrum from a model of simulated string networks first introduced by
Albrecht et al. \cite{ABR97} and further developed in Ref.~\cite{PV99} and
in this work. Our calculations indicate that the bispectrum resulting from
this model is negligible when compared with the cosmic variance. We discuss the implications of this result for detectability of cosmic strings through the bispectrum statistic.

\section{CMB Bispectrum from active models}
\label{sec:bispectrum}

We assume that, given a model of active perturbations, such as a string
simulation, we can calculate the energy-momentum tensor $T_{\mu \nu }({\mathbf
x},\tau )$ for a particular realization of the sources in a finite spatial
volume $V_{0}$.
Here, ${\bf x}$ is a 3-dimensional
coordinate and $\tau $ is the cosmic time.
Many simulations are run to obtain an ensemble of random
realizations of sources with statistical properties appropriate for the
given model. The spatial Fourier decomposition 
of $T_{\mu \nu }$ can be written as
\begin{equation} \label{fouriersum}
T_{\mu \nu }({{\mathbf x}},\tau )=\sum _{{\mathbf k}}\Theta_{\mu \nu
}({{\mathbf k}},\tau )e^{i{{\mathbf k}}{{\mathbf x}}}\, \, \, ,
\end{equation}
where ${{\mathbf k}}$ are discrete. If $V_{0}$ is sufficiently large
we can approximate the summation by the integral \begin{eqnarray}
\sum _{{\mathbf k}}\Theta_{\mu \nu }({{\mathbf k}},\tau )e^{i{{\mathbf
k}}{{\mathbf x}}}\approx \frac{V_{0}}{(2\pi )^{3}}\int d^{3}{{\mathbf
k}}\Theta_{\mu \nu }({{\mathbf k}},\tau )e^{i{{\mathbf k}}{{\mathbf x}}}\, \,
\, ,\label{sumtoint}
\end{eqnarray}
and the corresponding inverse Fourier transform will be \begin{equation}
\label{inversefourier}
\Theta_{\mu \nu }({{\mathbf k}},\tau )=\frac{1}{V_{0}}\int
_{V_{0}}d^{3}{{\mathbf x}}\,T_{\mu \nu }({{\mathbf x}},\tau )e^{-i{{\mathbf
k}}{{\mathbf x}}}\, \, \, .
\end{equation}
Of course, the final results, such as the CMB power spectrum or bispectrum,
do not depend on the choice of $V_{0}$. To ensure this independence, 
we shall keep $V_{0}$ in all expressions where it appears throughout the
following sections.

It is conventional to expand the temperature fluctuations over the basis of
spherical harmonics, \begin{equation}
{\Delta T/T}({\hat{{\mathbf n}}})=\sum
_{lm}a_{lm}Y_{lm}({\hat{{\mathbf n}}}),\end{equation}
where $\hat{{\mathbf n}}$ is a
unit vector. The coefficients $a_{lm}$ can be decomposed into Fourier
modes, \begin{equation} \label{eq:alm-def}
a_{lm}=\frac{V_{0}}{(2\pi )^{3}}\left( -i\right) ^{l}4\pi \int
d^{3}{{\mathbf k}}\, \Delta _{l}\left( {{\mathbf k}}\right)
Y^{*}_{lm}({\hat{{\mathbf k}}}).
\end{equation}
Given the sources $\Theta_{\mu \nu }({{\mathbf k}},\tau )$, the quantities
$\Delta _{l}({{\mathbf k}})$ are found by solving linearized
Einstein-Boltzmann equations and integrating along the line of sight, using
a code similar to CMBFAST \cite{cmbfast}. This standard procedure 
can be written
symbolically as the action of a linear operator ${\hat{B}}_{l}^{\mu \nu
}(k)$ on the source energy-momentum tensor, $\Delta _{l}({{\mathbf
k}})={\hat{B}}_{l}^{\mu \nu }(k)\Theta_{\mu \nu }({{\mathbf k}},\tau )$,
so the third moment of $\Delta _{l}({{\mathbf k}})$ is linearly related to the
three-point correlator of $\Theta_{\mu \nu }({{\mathbf k}},\tau )$. Below
we consider the quantities $\Delta _{l}({{\mathbf k}})$, corresponding to a
set of realizations of active sources, as given. The numerical procedure for computing $\Delta _{l}({{\mathbf k}})$ was developed in Refs.~\cite{ABR97} and \cite{PV99}.

The third moment of $a_{lm}$, namely $\left\langle
a_{l_{1}m_{1}}a_{l_{2}m_{2}}a_{l_{3}m_{3}}\right\rangle $, can be expressed
as \begin{eqnarray} &  & \left( -i\right) ^{l_{1}+l_{2}+l_{3}}\left( 4\pi
\right) ^{3}\! \! \frac{V_{0}^{3}}{(2\pi )^{9}}\! \! \int \! \!
d^{3}{{\mathbf k}}_{1}d^{3}{{\mathbf k}}_{2}d^{3}{{\mathbf
k}}_{3}Y^{*}_{l_{1}m_{1}}\! ({\hat{{\mathbf k}}}_{1})\nonumber \\ & \times
& Y^{*}_{l_{2}m_{2}}\! ({\hat{{\mathbf k}}}_{2})Y^{*}_{l_{3}m_{3}}\!
({\hat{{\mathbf k}}}_{3})\left\langle \Delta _{l_{1}}\! \! \left( {{\mathbf
k}}_{1}\right) \Delta _{l_{2}}\! \! \left( {{\mathbf k}}_{2}\right) \Delta
_{l_{3}}\! \! \left( {{\mathbf k}}_{3}\right) \right\rangle
.\label{eq:3alm-1}
\end{eqnarray}

A straightforward numerical evaluation of Eq.~(\ref{eq:3alm-1}) from given sources
$\Delta _{l}\left( {{\mathbf k}}\right) $ is prohibitively difficult,
because it involves too many integrations of oscillating functions.
However, we shall be able to reduce the computation to integrations over
scalars (a similar method was employed in \cite{KS00}). Due to homogeneity,
the 3-point function vanishes unless the triangle constraint is
satisfied,\begin{equation} \label{eq:triangle}
{{\mathbf k}}_{1}+{{\mathbf k}}_{2}+{{\mathbf k}}_{3}=0.
\end{equation}
We may write \begin{eqnarray}
&  & \left\langle \Delta _{l_{1}}\left( {{\mathbf k}}_{1}\right) \Delta
_{l_{2}}\left( {{\mathbf k}}_{2}\right) \Delta _{l_{3}}\left( {{\mathbf
k}}_{3}\right) \right\rangle \nonumber \\ & = & \delta ^{(3)}\left(
{{\mathbf k}}_{1}+{{\mathbf k}}_{2}+{{\mathbf k}}_{3}\right)
P_{l_{1}l_{2}l_{3}}\left( {{\mathbf k}}_{1},{{\mathbf k}}_{2},{{\mathbf
k}}_{3}\right) ,\label{p3lvector}
\end{eqnarray}
where the three-point function $P_{l_{1}l_{2}l_{3}}\left( {{\mathbf
k}}_{1},{{\mathbf k}}_{2},{{\mathbf k}}_{3}\right) $ is defined only for
values of ${\mathbf k}_{i}$ that satisfy Eq.~(\ref{eq:triangle}). Given the
scalar values $k_{1}$, $k_{2}$, $k_{3}$, there is a unique (up to an
overall rotation) triplet of directions ${\hat{{\mathbf k}}}_{i}$ for which
the RHS of Eq.~(\ref{p3lvector}) does not vanish. The quantity
$P_{l_{1}l_{2}l_{3}}\left( {{\mathbf k}}_{1},{{\mathbf k}}_{2},{{\mathbf
k}}_{3}\right) $ is invariant under an overall rotation of all three
vectors ${{\mathbf k}}_{i}$ and therefore may be equivalently represented
by a function of \emph{scalar} values $k_{1}$, $k_{2}$, $k_{3}$,
while preserving all angular information. Hence, we can rewrite Eq.~(\ref{p3lvector})
as \begin{eqnarray} &  & \left\langle \Delta _{l_{1}}\! \! \left( {{\mathbf
k}}_{1}\right) \Delta _{l_{2}}\! \! \left( {{\mathbf k}}_{2}\right) \Delta
_{l_{3}}\! \! \left( {{\mathbf k}}_{3}\right) \right\rangle \nonumber \\ &
= & \delta ^{(3)}\left( {{\mathbf k}}_{1}+{{\mathbf k}}_{2}+{{\mathbf
k}}_{3}\right) P_{l_{1}l_{2}l_{3}}(k_{1},k_{2},k_{3}).\label{p3lscalar}
\end{eqnarray}
Then, using the simulation volume $V_{0}$ explicitly, we have \begin{equation}
\label{p3l0}
P_{l_{1}l_{2}l_{3}}\! \left( k_{1},k_{2},k_{3}\right) \! =\! \frac{(2\pi
)^{3}}{V_{0}}\left\langle \Delta _{l_{1}}\! \! \left( {{\mathbf
k}}_{1}\right) \Delta _{l_{2}}\! \! \left( {{\mathbf k}}_{2}\right) \Delta
_{l_{3}}\! \! \left( {{\mathbf k}}_{3}\right) \right\rangle .
\end{equation}
Given an arbitrary direction $\hat{{\mathbf k}}_{1}$ and the magnitudes
$k_{1}$, $k_{2}$ and $k_{3}$, the directions $\hat{{\mathbf k}}_{2}$
and $\hat{{\mathbf k}}_{3}$ are specified up to overall rotations by the
triangle constraint. Therefore, both sides of Eq.~(\ref{p3l0}) are
functions of scalar $k_{i}$ only. The expression on the RHS of Eq.
(\ref{p3l0}) is evaluated numerically by averaging over different
realizations of the sources \textit{and} over permissible directions
$\hat{{\mathbf k}}_{i}$; below we shall give more details of the procedure.

Substituting Eqs.~(\ref{p3lscalar}) and (\ref{p3l0}) into (\ref{eq:3alm-1}),
Fourier transforming the Dirac delta and using the Rayleigh identity, we
can perform all angular integrations analytically and obtain a compact form
for the third moment, \begin{equation}
\label{eq:3alm-res}
\left\langle a_{l_{1}m_{1}}a_{l_{2}m_{2}}a_{l_{3}m_{3}}\right\rangle
={\mathcal{H}}_{l_{1}l_{2}l_{3}}^{m_{1}m_{2}m_{3}}\int r^{2}dr\,
b_{l_{1}l_{2}l_{3}}(r),
\end{equation}
where, denoting the Wigner $3j$-symbol by
$\left( ^{\, \, l_{1}\, \; l_{2}\, \; l_{3}}_{m_{1}m_{2}m_{3}}\right) $, we
have\begin{eqnarray} {\mathcal{H}}_{l_{1}l_{2}l_{3}}^{m_{1}m_{2}m_{3}} &
\equiv  & \sqrt{\frac{\left( 2l_{1}+1\right) \left( 2l_{2}+1\right) \left(
2l_{3}+1\right) }{4\pi }}\nonumber \\ &  & \times \left( \begin{array}{ccc}
l_{1} & l_{2} & l_{3}\\
0 & 0 & 0
\end{array}\right) \left( \begin{array}{ccc}
l_{1} & l_{2} & l_{3}\\
m_{1} & m_{2} & m_{3}
\end{array}\right) \, ,\label{eq:hlll}
\end{eqnarray}
and where we have defined the auxiliary quantities $b_{l_{1}l_{2}l_{3}}$
using spherical Bessel functions $j_{l}$, \begin{eqnarray}
b_{l_{1}l_{2}l_{3}}(r) & \equiv  & \frac{8}{\pi ^{3}}\frac{V_{0}^{3}}{(2\pi
)^{3}}\int k_{1}^{2}dk_{1}\, k_{2}^{2}dk_{2}\, k_{3}^{2}dk_{3}\, \nonumber
\\ & \times  &
j_{l_{1}}(k_{1}r)j_{l_{2}}(k_{2}r)j_{l_{3}}(k_{3}r)P_{l_{1}l_{2}l_{3}}(k_{1},k_{2},k_{3}).
\label{defineb}
\end{eqnarray}
The volume factor $V_{0}^{3}$ contained in this expression is correct: as
shown in the next section, each term $\Delta _{l}$ includes a factor
$V_{0}^{-2/3}$, while the average quantity
$P_{l_{1}l_{2}l_{3}}(k_{1},k_{2},k_{3})\propto V_{0}^{-3}$
{[}cf.~Eq.~(\ref{p3l0}){]}, so that the arbitrary volume $V_{0}$ of the
simulation cancels.

Our proposed numerical procedure therefore consists of computing the RHS of
Eq.~(\ref{eq:3alm-res}) by evaluating the necessary integrals. For fixed
$\left\{ l_{1}l_{2}l_{3}\right\} $, computation of the quantities
$b_{l_{1}l_{2}l_{3}}(r)$ is a triple integral over scalar $k_{i}$ defined
by Eq.~(\ref{defineb}); it is followed by a fourth scalar integral over $r$
{[}Eq.~(\ref{eq:3alm-res}){]}. We also need to average over many
realizations of sources to obtain $P_{l_{1}l_{2}l_{3}}\! \left(
k_{1},k_{2},k_{3}\right) $. It was not feasible for us to precompute the
values $P_{l_{1}l_{2}l_{3}}\! \left( k_{1},k_{2},k_{3}\right) $ on a grid
before integration because of the large volume of data: for each set
$\left\{ l_{1}l_{2}l_{3}\right\} $ the grid must contain $\sim 10^{3}$
points for each $k_{i}$. Instead, we precompute $\Delta _{l}\! \! \left(
{{\mathbf k}}\right) $ from one realization of sources and evaluate the RHS
of Eq.~(\ref{p3l0}) on that data as an \emph{estimator} of
$P_{l_{1}l_{2}l_{3}}\! \left( k_{1},k_{2},k_{3}\right) $, averaging over
allowed directions of $\hat{{\mathbf k}}_{i}$. The result is used for
integration in Eq.~(\ref{defineb}).

Because of isotropy and since the
allowed sets of directions $\hat{{\mathbf k}}_{i}$ are planar, it is enough
to restrict the numerical calculation to directions $\hat{{\mathbf k}}_{i}$
within a fixed two-dimensional plane. This significantly reduces the amount
of computations and data storage, since $\Delta _{l}\! \left( {{\mathbf
k}}\right) $ only needs to be stored on a two-dimensional grid of ${\mathbf
k}$.

In estimating $P_{l_{1}l_{2}l_{3}}\! \left( k_{1},k_{2},k_{3}\right) $
from Eq.~(\ref{p3l0}), averaging over directions of $\hat{{\mathbf k}}_{i}$
plays a similar role to ensemble averaging over source realizations.
Therefore if the number of directions is large enough (we used 720 for
cosmic strings), only a
moderate number of different source realizations is needed. The main
numerical difficulty is the highly oscillating nature of the function
$b_{l_{1}l_{2}l_{3}}(r)$. The calculation of the bispectrum for cosmic
strings presented in Section~\ref{sec:bisp} requires about
20 days of a single-CPU workstation time per realization.

We note that this method is specific for the bispectrum and cannot be applied
to compute higher-order correlations. The reason is that higher-order
correlations involve configurations of vectors ${\mathbf k}_{i}$ that are
not described by scalar values $k_{i}$ and not restricted to a plane. For
instance, a
computation of a 4-point function would involve integration of highly
oscillating functions over four vectors ${\mathbf k}_{i}$ which is
computationally infeasible.

From Eq.~(\ref{eq:3alm-res}) we derive the CMB angular bispectrum
${\mathcal{C}}_{l_{1}l_{2}l_{3}}$, defined as\cite{GM00}\begin{eqnarray}
\bigl \langle a_{l_{1}m_{1}}a_{l_{2}m_{2}}a_{l_{3}m_{3}}\bigr \rangle
=\left( \begin{array}{ccc} l_{1} & l_{2} & l_{3}\\
m_{1} & m_{2} & m_{3}
\end{array}\right) {\mathcal{C}}_{l_{1}l_{2}l_{3}}\, .
\end{eqnarray}
The presence of the 3$ j$-symbol guarantees that the third moment vanishes
unless $m_{1}+m_{2}+m_{3}=0$ and the $l_{i}$ indices satisfy the triangle
rule $|l_{i}-l_{j}|\leq l_{k}\leq l_{i}+l_{j}$. Invariance under spatial
inversions of the three-point correlation function implies the additional
`selection rule' $l_{1}+l_{2}+l_{3}=\mbox {even}$, in order for the third
moment not to vanish. Finally, from this last relation and
using standard properties of the 3$ j$-symbols, it follows that the
angular bispectrum ${\mathcal{C}}_{l_{1}l_{2}l_{3}}$ is left unchanged
under any arbitrary permutation of the indices $l_{i}$.

In this paper we restrict our calculations to the angular bispectrum
$C_{l_{1}l_{2}l_{3}}$ in the `diagonal' case, \textit{i.e.}~$l_{1}=l_{2}=l_{3}=l$.
This is a representative case and, in fact, the one most frequently
considered in the literature. Plots of the power spectrum are usually done
in terms of $l(l+1)C_{l}$ which, apart from constant factors, is the
contribution to the mean squared anisotropy of temperature fluctuations per
unit logarithmic interval of $l$. In full analogy with this, the relevant
quantity to work with in the case of the bispectrum is
\begin{eqnarray} G_{lll} = l(2l+1)^{3/2}\left( \begin{array}{ccc}
l & l & l\\
0 & 0 & 0
\end{array}\right) C_{lll}\, .\label{eq:QtP}
\end{eqnarray}
Details of this derivation are presented in the Appendix. For large values
of the multipole index $l$, $G_{lll}\propto l^{3/2}C_{lll}$.
Note also what happens with the 3$ j$-symbols appearing in the definition
of the coefficients ${\mathcal{H}}_{l_{1}l_{2}l_{3}}^{m_{1}m_{2}m_{3}}$:
the symbol $\left( ^{\, \, l_{1}\, \; l_{2}\, \;
l_{3}}_{m_{1}m_{2}m_{3}}\right) $ is absent from the definition of
$C_{l_{1}l_{2}l_{3}}$, while in Eq.~(\ref{eq:QtP}) the symbol $\left( ^{\,
l\; l\; l}_{0\: 0\: 0}\right) $ is squared. Hence, there are no remnant
oscillations due to the alternating sign of $\left( ^{\, l\; l\; l}_{0\:
0\: 0}\right) $.

However, even more important than the value of $C_{lll}$ itself is the
relation between the bispectrum and the cosmic variance associated
with it.
In fact, it is their comparison that tells us about the observability `in
principle' of the non-Gaussian signal. The cosmic variance constitutes a
theoretical uncertainty for all observable quantities and comes about due
to the fact of having just one realization of the stochastic process, in
our case, the CMB sky \cite{sc91}.

The way to proceed is to employ an estimator $\hat{C}_{l_{1}l_{2}l_{3}}$
for the bispectrum and compute the variance from it. By choosing an
unbiased estimator we ensure it satisfies $C_{l_{1}l_{2}l_{3}}=\langle
\hat{C}_{l_{1}l_{2}l_{3}}\rangle $. However, this condition does not
isolate a unique estimator. The proper way to select the {\it best unbiased}
estimator is to compute the variances of all candidates and choose
the one with the smallest value.
The estimator with this property was computed in Ref.~\cite{GM00}
and is \begin{equation} \label{eq:clll-best}
\hat{C}_{l_{1}l_{2}l_{3}}=\! \! \! \sum _{m_{1},m_{2},m_{3}}\! \! \left(
\begin{array}{ccc} l_{1} & l_{2} & l_{3}\\
m_{1} & m_{2} & m_{3}
\end{array}\right) a_{l_{1}m_{1}}a_{l_{2}m_{2}}a_{l_{3}m_{3}}.
\end{equation}
The variance of this estimator, assuming a mildly non-Gaussian
distribution, can be expressed in terms of the angular power spectrum
$C_{l}$ as follows~\cite{GM99}
\begin{equation}
\label{eq:sigma}
 \sigma
^{2}_{\hat{C}_{l_{1}l_{2}l_{3}}}\! \! \! \! =C_{l_{1}}C_{l_{2}}C_{l_{3}}\!
\left( 1\! +\! \delta _{l_{1}l_{2}}\! \! +\! \delta _{l_{2}l_{3}}\! \! +\!
\delta _{l_{3}l_{1}}\! \! +\! 2\delta _{l_{1}l_{2}}\delta
_{l_{2}l_{3}}\right) .
\end{equation}
The theoretical signal-to-noise ratio for the bispectrum is then given
by
\begin{equation}
(S/N)_{l_{1}l_{2}l_{3}} = 
|C_{l_{1}l_{2}l_{3}}/\sigma_{\hat{C}_{l_{1}l_{2}l_{3}}}|.
\end{equation}
In turn, for the diagonal case $l_{1}=l_{2}=l_{3}=l$ we have
\begin{equation} 
(S/N)_{l} = |C_{lll}/\sigma _{\hat{C}_{lll}}|.
\end{equation} 

Incorporating all the specifics of the particular experiment, such as sky
coverage, angular resolution, etc., will allow us to give an estimate of the
particular non-Gaussian signature associated with a given active source and, if
observable, indicate the appropriate range of multipole $l$'s where it is best
to look for it.

\section{Bispectrum from strings}
\label{sec:bisp}
\subsection{The string model}

To calculate the sources of perturbations we use an updated version of
the cosmic string model first introduced by Albrecht et al. \cite{ABR97} and
further developed in Ref.~\cite{PV99}, where the wiggly nature of
strings was taken into account. In these previous works the model was
tailored to the computation of the two-point statistics
(matter and CMB power spectra). When dealing with
higher-order statistics, such as the bispectrum, a different strategy
needs to be employed.

In the model, the string network is represented by a collection of uncorrelated
straight string segments produced at some early epoch and moving with random
uncorrelated velocities. At every subsequent epoch, a certain fraction of the
number of segments decays in a way that maintains network scaling. The length
of each segment at any time is taken to be equal to the correlation length of
the network. This and the root mean square velocity of segments are computed
from the velocity-dependent one-scale model of Martins and Shellard \cite{MS}.
The positions of segments are drawn from a uniform distribution in space, and
their orientations are chosen from a uniform distribution on a two-sphere.

The total energy of the string network in a volume $V$ at any time is
$E=N\mu L$, where $N$ is the total number of string segments at that time,
$\mu $ is the mass per unit length, and $L$ is the length of one segment. If
$L$ is the correlation length of the string network then, according to the
one-scale model, the energy density is $\rho ={E/V}={\mu /L^{2}}$, where
$V=V_{0}a^{3}$, the expansion factor $a$ is normalized so that $a=1$ today,
and $V_{0}$ is a constant simulation volume. It follows that
$N=V/L^{3}=V_{0}/\ell^{3}$, where $\ell=L/a$ is the comoving correlation length.
In the scaling regime $\ell$ is approximately proportional to the conformal
time $\tau $ and so the number of strings $N(\tau )$ within the simulation
volume $V_{0}$ falls as $\tau ^{-3}$.

To calculate the CMB anisotropy one
needs to evolve the string network over at least four orders of magnitude
in cosmic expansion. Hence, one would have to start with $N\gtrsim 10^{12}$
string segments in order to have one segment left at the present time.
Keeping track of such a huge number of segments is numerically infeasible.
A way around this difficulty was suggested in
Ref.\cite{ABR97}, where the idea was to consolidate all string segments
that decay at the same epoch. The number of segments that decay by the
(discretized) conformal time $\tau _{i}$ is \begin{equation} \label{eq:nd}
N_{d}(\tau _{i})=V_{0}\left( n(\tau _{i-1})-n(\tau _{i})\right) ,
\end{equation}
where $n(\tau )=[\ell(\tau )]^{-3}$ is the number density of strings at time
$\tau $. The energy-momentum tensor in Fourier space, $\Theta^{i}_{\mu \nu }$,
of these $N_{d}(\tau _{i})$ segments is a sum \begin{equation}
\label{emtsum}
\Theta^{i}_{\mu \nu }=\sum _{m=1}^{N_{d}(\tau _{i})}\Theta^{im}_{\mu \nu }\,
\, \, , \end{equation}
where $\Theta^{im}_{\mu \nu }$ is the Fourier transform of the energy-momentum
of the $m$-th segment. If segments are uncorrelated, then
\begin{equation}
\label{eq:theta2}
\langle\Theta^{im}_{\mu\nu}\Theta^{im'}_{\sigma\rho}\rangle =
\delta_{m m'} \langle\Theta^{im}_{\mu\nu}\Theta^{im}_{\sigma\rho}\rangle
\end{equation}
and
\begin{equation}
\langle\Theta^{im}_{\mu\nu}\Theta^{im'}_{\sigma\rho}
\Theta^{im''}_{\gamma\delta}\rangle =
\delta_{m m'}\delta_{m m''}
\langle\Theta^{im}_{\mu\nu}\Theta^{im}_{\sigma\rho}
\Theta^{im}_{\gamma\delta}\rangle .
\end{equation}
Here the angular brackets $\langle \ldots \rangle $ denote the
ensemble average, which in our case means averaging over many realizations
of the string network. If we are calculating power spectra, then the
relevant quantities are the two-point functions of $\Theta^{i}_{\mu \nu }$,
namely
\begin{eqnarray} \langle \Theta^{i}_{\mu \nu }\Theta^{i}_{\sigma \rho
}\rangle =\langle \sum _{m=1}^{N_{d}(\tau _{i})}\sum _{m'=1}^{N_{d}(\tau
_{i})}\Theta^{im}_{\mu \nu }\Theta^{im'}_{\sigma \rho }\rangle
.\label{thefix1} \end{eqnarray}
Eq.~(\ref{eq:theta2}) allows us to write \begin{eqnarray}
\langle \Theta^{i}_{\mu \nu }\Theta^{i}_{\sigma \rho }\rangle =\sum
_{m=1}^{N_{d}(\tau _{i})}\langle \Theta^{im}_{\mu \nu }\Theta^{im}_{\sigma
\rho }\rangle =N_{d}(\tau _{i})\langle \Theta^{i1}_{\mu \nu
}\Theta^{i1}_{\sigma \rho }\rangle ,\label{thefix2}
\end{eqnarray}
where $\Theta^{i1}_{\mu \nu }$ is of the energy-momentum
of one of the segments that decay by the time $\tau _{i}$. The last step in
Eq.~(\ref{thefix2}) is possible because the segments are statistically
equivalent. Thus, if we only want to reproduce the correct power spectra in
the limit of a large number of realizations, we can replace the sum in
Eq.~(\ref{emtsum}) by \begin{equation} \label{thefix3}
\Theta^{i}_{\mu \nu }=\sqrt{N_{d}(\tau _{i})}\Theta^{i1}_{\mu \nu }.
\end{equation}
The total energy-momentum tensor of the network in Fourier space is a sum over
the consolidated segments:
\begin{equation}
\label{emtsum1}
\Theta_{\mu \nu }=\sum _{i=1}^{K}\Theta^{i}_{\mu \nu }=\sum
_{i=1}^{K}\sqrt{N_{d}(\tau _{i})}\Theta^{i1}_{\mu \nu }\, .
\end{equation}
So, instead of summing over $\sum _{i=1}^{K}N_{d}(\tau _{i})\gtrsim 10^{12}$
segments we now sum over only $K$ segments, making $K$ a parameter.

For the three-point functions we extend the above procedure. Instead of
Eqs.~(\ref{thefix1}) and (\ref{thefix2}) we now write
\begin{eqnarray}
\langle \Theta^{i}_{\mu \nu }\Theta^{i}_{\sigma \rho }\Theta^{i}_{\gamma
\delta }\rangle =\langle \sum _{m=1}^{N_{d}(\tau _{i})}\sum
_{m'=1}^{N_{d}(\tau _{i})}\sum _{m''=1}^{N_{d}(\tau _{i})}\Theta^{im}_{\mu \nu
}\Theta^{im'}_{\sigma \rho }\Theta^{im''}_{\gamma \delta }\rangle  &  &
\nonumber \\ =\sum _{m=1}^{N_{d}(\tau _{i})}\langle \Theta^{im}_{\mu \nu
}\Theta^{im}_{\sigma \rho }\Theta^{im}_{\gamma \delta }\rangle =N_{d}(\tau
_{i})\langle \Theta^{i1}_{\mu \nu }\Theta^{i1}_{\sigma \rho
}\Theta^{i1}_{\gamma \delta }\rangle \, .\label{newfix} \end{eqnarray}
Therefore, for the purpose of calculation of three-point functions, the sum
in Eq.~(\ref{emtsum}) should now be replaced by \begin{equation}
\label{newfix1}
\Theta^{i}_{\mu \nu }=[N_{d}(\tau _{i})]^{1/3}\Theta^{i1}_{\mu \nu }\, .
\end{equation}

Both expressions in Eqs.~(\ref{thefix3}) and (\ref{newfix1}), depend on the
simulation volume, $V_{0}$, contained in the definition of $N_{d}(\tau _{i})$
given in Eq.~(\ref{eq:nd}). This is to be expected and is consistent with our
calculations, since this volume cancels in expressions for observable
quantities.

Note also that the simulation model in its present form does not allow
computation of CMB sky maps. This is because the method of finding the two- and
three-point functions as we described involves {}``consolidated{}'' quantities
$\Theta^{i}_{\mu \nu }$ which do not correspond to the energy-momentum tensor
of a real string network. These quantities are auxiliary and specially prepared
to give the correct two- or three-point functions after ensemble averaging.

\subsection{Results and discussion}

In Fig.~\ref{fig:1} we show the results for
$G_{lll}^{1/3}$ {[}cf.~Eq.~(\ref{eq:QtP}){]}. It was calculated using
the string model with $800$ consolidated segments in a
flat universe with cold dark matter and a cosmological constant. Only
the scalar contribution to the anisotropy has been included. Vector and
tensor contributions are known to be relatively insignificant for local
cosmic strings and can safely be ignored in this model
\cite{ABR97,PV99}\footnote{The contribution of vector and tensor modes is large in
the case of global strings \cite{TPS98,dgs96}.}.
The plots are produced using a single realization of
the string network by averaging over $720$ directions of ${\mathbf
k}_{i}$. The comparison of $G_{lll}^{1/3}$ (or equivalently
${C}_{lll}^{1/3}$) with its cosmic variance {[}cf.~Eq.~(\ref{eq:sigma}){]} clearly shows that the bispectrum (as computed
from our cosmic string model) lies hidden in the theoretical noise and
is therefore undetectable for any given value of $l$.

Let us note, however, that in its present
stage our string code describes Brownian, wiggly long strings in spite
of the fact that long strings are very likely not Brownian on the
smallest scales \cite{Martins}. In addition, the presence of small
string loops \cite{proty} and gravitational radiation into which they decay were not
yet included in our model. These are important effects that could, in
principle, change our predictions for the string-generated CMB
bispectrum on very small angular scales.

\begin{figure}
{\par\centering \leavevmode\epsfig{file=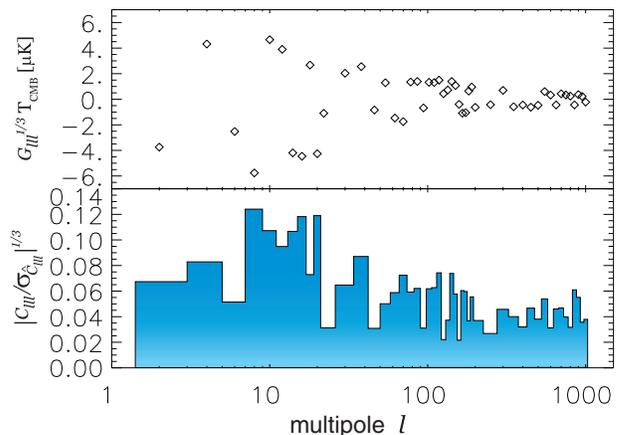, width=9cm} \par}
\caption{The CMB angular bispectrum in the `diagonal' case ($
G_{lll}^{1/3}$) from wiggly cosmic strings in a spatially flat model
with cosmological parameters $ \Omega _{\textrm{CDM}}=0.3$, $ \Omega
_{\textrm{baryon}}=0.05$, $ \Omega _{\Lambda }=0.65$, and Hubble
constant $ H=0.65{\textrm{km}}{\textrm{s}}^{-1}{\textrm{Mpc}}^{-1}$
{[}upper panel{]}. In the lower panel we show the ratio of the signal
to theoretical noise $ |C_{lll}/\sigma _{\hat{C}_{lll}}|^{1/3}$ for
different multipole indices.
Normalization follows from fitting the power spectrum to the
BOOMERanG and MAXIMA data.}
\label{fig:1}
\end{figure}

The imprint of cosmic strings on the CMB is a combination of different
effects. Prior to the time of recombination strings induce density and
velocity fluctuations on the surrounding matter. During the period of last
scattering these fluctuations are imprinted on the CMB through the Sachs-Wolfe
effect: namely, temperature fluctuations arise because relic photons encounter a
gravitational potential with spatially dependent depth. In addition to
the Sachs-Wolfe effect, moving long strings drag the surrounding plasma and
produce velocity fields that cause temperature anisotropies due to
Doppler shifts. While a string segment by itself is a highly non-Gaussian
object, fluctuations induced by string segments before recombination
are a superposition of effects of many random strings stirring the
primordial plasma. These fluctuations are thus expected to be Gaussian as a
result of the central limit theorem.

As the universe becomes transparent, strings continue to leave their
imprint on the CMB mainly due to the Kaiser-Stebbins effect \cite{KS84}. This
effect results in line discontinuities in the temperature field of photons
passing on opposite sides of a moving long string.\footnote{In an
extension of the Kaiser-Stebbins effect, Benabed and Bernardeau
\cite{coucou} have recently considered the generation of a B-type polarization field
out of E-type polarization, through gravitational lensing on a cosmic string.}
However, this effect can result in non-Gaussian perturbations only on
sufficiently small scales. This is because on scales larger than the characteristic
inter-string separation at the time of the radiation-matter equality,
the CMB temperature
perturbations result from superposition of effects of many strings and are
likely to be Gaussian. Avelino \textit{et al.} \cite{ASWA98} applied several
non-Gaussian tests to the perturbations seeded by cosmic strings. They found
the density field distribution to be close to Gaussian on scales larger than
$1.5 (\Omega _M h^2)^{-1}$ Mpc, where $\Omega _M$ is the
fraction of cosmological matter density in baryons and CDM combined. Scales
this small correspond to the multipole index of order $l \sim 10^4$. We have
not attempted a calculation the CMB bispectrum on these scales because the 
linear approximation is almost guaranteed to fail at such small scales, 
and because of increased computational cost for higher $l$ multipoles.

In summary, we have developed a numerical method to compute from first
principles one of the cleanest non-Gaussian discriminators---the CMB
angular bispectrum---in any active model of structure formation, such
as cosmic defects, where the energy-momentum tensor is known
or can be simulated. Our method does not use CMB sky maps
and requires a moderate amount of computations. We applied this
method to the computation of some relevant components of the bispectrum
produced from a model of cosmic strings and found that the non-zero
non-Gaussian signal is unobservable even with forthcoming satellite-based CMB
missions. Further computations and improvements using this method will
be reported elsewhere.

\section*{Acknowledgments}

L.P. and S.W. give special thanks to Tanmay Vachaspati for valuable
insights.
A.G. thanks I.A.P. and D.A.R.C. for hospitality in Paris, France, and
CONICET and Fundaci\'on Antorchas for financial support.

\appendix

\section{Plotting the bispectrum}

We explain here our choice for the normalization of the
angular bispectrum, as given in Eq.~(\ref{eq:QtP}). The argument
is an extension of the case of the angular power spectrum.

We can express the two-point correlation function at zero lag in terms
of the angular spectrum as follows
\begin{equation} \left\langle \left(
\frac{\Delta T}{T}\right) ^{2}\right\rangle =\frac{1}{4\pi }\sum
_{l}(2l+1)C_{l}.
\end{equation}
In the small angular scale limit, the approximation that $C_{l}$ is a
smooth function of the multipole index $l$ is well justified. We can then
replace the sum by an integral and get \begin{equation}
\left\langle \left( \frac{\Delta T}{T}\right) ^{2}\right\rangle \approx
\frac{1}{4\pi }\int \frac{dl}{l}l(2l+1)C_{l}.
\end{equation}
Now, ${dl/l}=d(\ln (l))$ and therefore $l(2l+1)C_{l}$ is the contribution
to the mean squared anisotropy of temperature fluctuations per unit
logarithmic interval of $l$. In standard practice, one usually plots
$l(l+1)C_{l}/2\pi$ versus $l$, which for large $l$ is proportional to
$l(2l+1)C_{l}/4\pi$. On small angular scales, then, this is $\propto l^{2}C_{l}$.

In the case of the three-point correlation function the situation is a bit
more involved. Let us consider the skewness:
\begin{eqnarray} && \left\langle \left(
\frac{\Delta T}{T}\right)^{3}\right\rangle =\sum _{l_{1}l_{2}l_{3}}\sqrt{\frac{2l_{1}+1}{4\pi}}
\nonumber \\ &&\times \sqrt{\frac{2l_{2}+1}{4\pi }}\sqrt{\frac{2l_{3}+1}{4\pi }}\left(
\begin{array}{ccc} l_{1} & l_{2} & l_{3}\\
0 & 0 & 0
\end{array}\right) {\mathcal{C}}_{l_{1}l_{2}l_{3}}\label{ccc3}
\end{eqnarray}
We know that ${\mathcal{C}}_{l_{1}l_{2}l_{3}}$
is smooth in all three indices. We can split the skewness into three sums:
the sum of terms where all $l_{i}$ are different, the sum where only two of
the three $l_{i}$ are different, and the sum of terms where all $l_{i}$
are equal. Omitting constant factors of $4\pi $, we outline the same
procedure as above for the two-point function. For the first sum we get
\begin{eqnarray} \! \! \int \! \! \frac{dl_{1}}{l_{1}}\! \! \int \! \!
\frac{dl_{2}}{l_{2}}\! \! \int \! \! \frac{dl_{3}}{l_{3}} &  & \! \! \! \!
l_{1}\sqrt{2l_{1}\! \! +\! \! 1} \,\, l_{2}\sqrt{2l_{2}\! \! +\! \!
1} \,\, l_{3}\sqrt{2l_{3}\! \! +\! \! 1}\nonumber \\ & \times  & \left(
\begin{array}{ccc} l_{1} & l_{2} & l_{3}\\
0 & 0 & 0
\end{array}\right) C_{l_{1}l_{2}l_{3}}\, ,
\end{eqnarray}
while for the second sum we have \begin{eqnarray}
&  & \! \! \int \! \! \frac{dl_{1}}{l_{1}}\! \! \int \! \!
\frac{dl_{2}}{l_{2}}l_{1}\sqrt{2l_{1}\! \! +\! \! 1} \,\, l_{2}\left( 2l_{2}\! \!
+\! \! 1\right) \left( \begin{array}{ccc} l_{1} & l_{2} & l_{2}\\
0 & 0 & 0
\end{array}\right) C_{l_{1}l_{2}l_{2}}\, ,\nonumber \\
&  &
\end{eqnarray}
and for the third \begin{eqnarray}
\int \frac{dl}{l}l(2l+1)^{3/2}\left( \begin{array}{ccc}
l & l & l\\
0 & 0 & 0
\end{array}\right) C_{lll}. &
\end{eqnarray}
If one is interested in the diagonal terms $C_{lll}$, then,
following the last equation, the relevant quantity to plot is given by
\begin{eqnarray} l(2l+1)^{3/2}\left( \begin{array}{ccc}
l & l & l\\
0 & 0 & 0
\end{array}\right) C_{lll}\, ,
\end{eqnarray}
which is $\propto l^{3/2}C_{lll}$ at large $l$.


\begin{thebibliography}{10}
\bibitem{cmblevon}The numerical code employed here will be made publicly available shortly.
\end{thebibliography}
\end{document}